\documentclass[openacc]{rsproca_new}

\usepackage[normalem]{ulem}

\begin{document}

\title{Globular Cluster Formation and Evolution in the Context of Cosmological Galaxy Assembly: Open Questions}

\author{
Duncan A. Forbes$^{1}$, Nate Bastian$^{2}$, Mark Gieles$^{3}$, Robert A. Crain$^{2}$, J.~M.~Diederik Kruijssen$^4$, S\o ren~S.~Larsen$^{5}$, Sylvia Ploeckinger$^{6}$, Oscar Agertz$^{7}$, Michele Trenti$^{8}$, Annette M. N. Ferguson$^{9}$, Joel Pfeffer$^{2}$, Oleg Y. Gnedin$^{10}$}

\address{$^{1}$Centre for Astrophysics \& Supercomputing, Swinburne University, Hawthorn VIC 3122, Australia\\
$^{2}$Astrophysics Research Institute, Liverpool John Moores University,
146 Brownlow Hill, Liverpool L3 5RF, United Kingdom\\
$^{3}$Department of Physics, University of Surrey, Guildford GU2 7XH, United Kingdom\\
$^{4}$Astronomisches Rechen-Institut, Zentrum f\"{u}r Astronomie der Universit\"{a}t Heidelberg, Monchhofstra\ss e 12-14, 69120 Heidelberg, Germany \\
$^{5}$Department of Astrophysics/IMAPP, Radboud University, PO Box 9010, 6500 GL Nijmegen, The Netherlands\\
$^{6}$Leiden Observatory, Leiden University, PO Box 9513, NL-2300 RA Leiden, the Netherlands\\
$^{7}$Lund Observatory, Department of Astronomy \& Theoretical Physics, Lund University, Box 43, SE-221 00 Lund, Sweden \\
$^{8}$School of Physics, The University of Melbourne, VIC, 3010, Australia \\ 
$^{9}$Institute for Astronomy, University of Edinburgh, Blackford Hill, Edinburgh EH9 3HJ, United Kingdom \\
$^{10}$Department of Astronomy, University of Michigan, Ann Arbor, MI, 48109, USA
}

\subject{astrophysics}


\corres{Duncan Forbes\\
\email{dforbes@swin.edu.au}}




\maketitle

\noindent
{\bf ABSTRACT}\\
We discuss some of the key open questions regarding the formation and evolution of globular clusters (GCs) during galaxy formation and assembly within a cosmological framework. The current state-of-the-art for both observations and simulations is described, and we briefly mention directions for future research.
The oldest GCs have ages $\ge$ 12.5 Gyr and formed around the time of reionisation. Resolved colour-magnitude diagrams of Milky Way GCs and direct imaging of lensed proto-GCs at z $\sim$ 6 with JWST promise further insight. Globular clusters are known to host multiple populations of stars with variations in their chemical abundances. Recently, such multiple populations have been detected in $\sim$2 Gyr old compact, massive star clusters. This suggests a common, single pathway for the formation of GCs at high and low redshift. The shape of the initial mass function for GCs remains unknown, however for massive galaxies a power-law mass function is favoured. Significant progress has been made recently modelling GC formation in the context of galaxy formation, with success in reproducing many of the observed GC-galaxy scaling relations. 

\section{Preamble}

A small group of researchers were invited by the Royal Society to attend a workshop at Chicheley Hall, Buckinghamshire.
Over the course of 2 days (5, 6 April 2017) they presented new results and discussed globular cluster formation and evolution in the context of galaxy assembly. The article that follows represents some of the discussion from that meeting with a focus on the open questions that were raised and the prospects for answering those questions in the near future. 
We hope that this article will be of value to other researchers in this field, as well as those in related areas of research.

\section{When did globular clusters form?}
%
%
\label{sec:when_gcs_form}

Knowing the age of globular clusters (GCs), and hence their redshift of formation for a given cosmology, is a key parameter for understanding their relationship to galaxy formation and for testing model predictions. For example, did GCs form more than 12.8 Gyr ago, i.e. before the end of reionisation (z $\approx$ 6) 
and perhaps play a role in reionising the universe, or did they mostly form at later times closer to the peak in the cosmic star formation rate (z $\approx$ 2)? Coupled with GC metallicities, knowing the age of GCs provides a stringent constraint on models of GC formation (e.g.  Muratov \& Gnedin 2010;  Li \& Gnedin 2014; Renaud, Agertz \& Gieles 2017; Kruijssen et al.~2018).

Almost seventy Milky Way GCs have age measurements based on deep colour-magnitude diagrams (CMDs) observed with the ACS onboard the Hubble Space Telescope.  The  resulting  age-metallicity relation reveals a  dominant population of  very old GCs covering a wide range of metallicities and a `young' branch for which age and metallicity are anti-correlated (Marin-Franch et al. 2009; Forbes \& Bridges 2010; VandenBerg et al. 2013; Leaman et al. 2013).  These studies generally find the very old metal-poor (MP) subpopulation to be somewhat older than the metal-rich (MR) subpopulation (i.e. $\sim$12.5 vs $\sim$11.5 Gyr respectively), but coeval ages can not yet be ruled out within the uncertainties. 
The `young branch' GCs can be largely associated with the  disrupted Sgr dwarf and other possible accreted GCs. We note that only a handful of Milky Way GCs have their ages determined from the white dwarf cooling sequence (e.g. Hansen et al. 2013; Garcia-Berro et al. 2014) which gives ages consistent with those measured from main sequence fitting. The method has the advantage of being less sensitive to metallicity but it requires very deep CMDs. 

What are the prospects for getting more  age measurements of Milky Way GCs? It would be very useful to increase the current sample of HST-observed Milky Way GCs, particularly metal-rich bulge GCs which are deficient in the current sample, largely due to foreground reddening issues.
A new technique, which will help to address this issue, is to derive the absolute ages of GCs based on deep near-infrared CMDs that probe the main sequence `kink'.  In the near-infrared, the CMDs of old stellar populations display a ``kink'' at $\sim0.5~M_{\odot}$ (corresponding to J--K $\sim$ 0.8) due to opacity effects of H$_2$.  The magnitude difference between the main sequence turn-off and this kink is sensitive to age, and largely insensitive to uncertainties in the extinction and distance modulus as well as uncertainties in the stellar models (Bono et al.~2010).  The location of the kink is also dependent on the metallicity, but there are hopes that with precise JWST photometry, absolute ages will be possible to slightly less than 1~Gyr accuracy (e.g. Correnti et al.~2016; Saracino et al. 2016).  It will also be interesting to determine relative ages for GCs with similar metallicities, which will provide a strong constraint on the timescale over which (at fixed metallicity) the GC system of the Milky Way was built up. 
An exciting possibility for future work with JWST is to measure the brown dwarf cooling sequence (Caiazzo et al. 2017).  Although challenging, this method has the advantage of being insensitive to distance (a key element in the current error budget for GC age determinations). 

We note that recently, six old LMC GCs were observed using ACS by Wagner-Kaiser et al. (2017).  They find relative ages that are coeval with the old halo GCs of the Milky Way as measured by Marin-Franch et al. (2009). 

Age estimates for GCs beyond the Local Group come from integrated light. Spectra with moderate signal-to-noise were obtained in the early 2000s with LRIS on the Keck 10m (e.g. Brodie et al. 2005), GMOS on Gemini 8m (e.g. Pierce et al. 2006) and FORS2 on the VLT 8m (e.g. Puzia et al. 2005). The situation was summarised in a meta-analysis in 2005 by Strader et al. (2005). They concluded that both MP and MR subpopulations were {\it ``no younger than their Galactic counterparts, with ages $\ge$ 10 Gyr''}. The absorption line index measurements from these spectra had values lower than predicted by 14 Gyr old stellar population tracks (i.e. formally indicating ages greater than the age of the universe). This inconsistency prevented the derivation of {\it absolute} ages. On the other hand measurement uncertainties prevented the detection of {\it relative} age differences of less than 1--2 Gyr. The observations have a bias to targeting brighter (more massive) GCs and those located in galaxy central regions which have a higher number density. Determining the relative ages of inner and outer (i.e. predominately in-situ and accreted - see \S~3) GCs is desirable. In the last decade, there have been very few new attempts to further measure the ages of GCs from spectra in galaxies beyond the Local Group. 

An alternative method for estimating GC ages was presented by Forbes et al. (2015), using measurements of GC mean metallicities and assuming that they followed a galaxy mass-metallicity relation (Kruijssen 2014)  extrapolated in redshift (and beyond the current observations). With these caveats in mind, which may lead to systematic uncertainties of $\ge$1 Gyr, the method indicates MR GC mean ages of 11.5 Gyr (z$_{\mathrm{form}}$ = 2.9) and MP ages of 12.5--12.8 Gyr (z$_{\mathrm{form}}$ = 4.8--5.9). If correct, this would place GC formation around the time of reionisation, and continuing after the epoch of reionisation. The age (epoch of formation) and predictions from various GC simulations pre-2015 are summarised in Forbes et al. (2015) -- none can be convincingly  ruled out on the basis of current GC age measurements. 

To directly measure mean relative ages for large numbers of extragalactic GCs, requires the multiplexing advantages of a very wide-field spectrograph that operates at blue wavelengths on a large telescope. A spectrograph that meets these requirements is the Prime Focus Spectrograph (PFS) to be commissioned on the Subaru 8.2m telescope in 2019. PFS has 2400 fibers that can be allocated over a 1.3 degree diameter field-of-view. It operates from 0.38 to 1.26 microns, thus covering the key absorption sensitive to age in old stellar populations. PFS will be able to obtain spectra for GCs in the inner and outer regions of nearby galaxies, and for several group/cluster galaxies in a single pointing. Careful background subtraction of the host galaxy and sky will be required. In addition, measuring mean relative ages from these spectra will require, at the very least, high S/N and the reduction of random errors from a large number of objects. 
Even for the nearest extragalactic GCs, this work will be very challenging on current 8-10m class telescopes 
and is better suited to the new 20-40m class telescopes. On the theoretical side, improvements are also needed in the modelling of integrated-light age diagnostics, e.g. the contribution of blue horizontal branch stars, thermally pulsing red giant branch stars, etc. The expansion of the stellar library to lower metallicity, as carried out recently by Villaume et al. (2017), is also particularly important for application of the models to GCs. 

As noted recently by Renzini (2017), we are on the cusp of observing the formation of GCs directly at high redshift. Pioneering work by Johnson,~T.~L. et al. (2017) and Vanzella et al. (2017) 
has utilized strong lensing amplification in distant (z $\sim$ 2--6) galaxy clusters imaged by HST to measure a small number of compact objects with typical sizes of 16--140 pc and masses of a few $\times$ 10$^6$ M$_{\odot}$. Even if such sizes still greatly exceed those of average GCs observed in the local Universe, these objects have been interpreted as GCs in formation. For one source at $z = 3.1$, Vanzella et al. (2017) measured a dynamical mass consistent with its stellar mass, implying no evidence of a mini dark matter halo. For an object at $z = 6.1$, they measured a size ($\sim$20 pc) and mass ($\sim$10$^7$ M$_{\odot}$) consistent with that expected for a proto-GC. Spectroscopy and imaging for large numbers of such compact objects at high redshift should be possible with JWST and help to determine if they are indeed proto-GCs. 

\section{Where did globular clusters form?}
\label{sec:formlocation}
Current ideas for massive galaxy formation focus on the concept of two-phase galaxy formation.
Cosmological simulations 
(e.g. Oser et al. 2010; Cook et al. 2016; Rodriguez-Gomez et al. 2016) 
indicate that large galaxies underwent an initial {\it in-situ} phase of formation followed by mass growth via an {\it ex-situ} (or accretion) phase. In these simulations star formation can be tagged as occuring within the potential well/virial radius of the host primary galaxy (i.e. {\it in-situ}) or within the satellite galaxy that is eventually accreted (i.e. {\it ex-situ}). Thus in the simulations, {\it in-situ} and {\it ex-situ} formed material can be defined and tracked separately.
These simulations predict that the  most massive galaxies  have accreted a large fraction of their final mass, while for the low mass galaxies their assembly is dominated by {\it in-situ} star formation. 
This suggests that today's galaxies should contain contributions from both {\it in-situ} and {\it ex-situ} formed GCs depending on the galaxy's assembly history.

The distinction between {\it in-situ} formed GCs and those that were accreted from another galaxy, is much harder to discern observationally. Indeed the  
concept itself is less well  defined during the initial phase of galaxy formation at high redshift (which may involve clumpy dissipative collapse, and inflowing cold gas onto a chaotic, turbulent gas disk) and at later times when gas is involved (e.g. GCs may form in the primary galaxy from low-metallicity gas that was acquired from a satellite).
Bearing these caveats in mind, an open question  within the context of current galaxy formation models is: where did GCs form?


The stellar components of  massive early-type galaxies and their red GCs have many properties in common (e.g. Pota et al. 2013). Thus it is often assumed that the metal-rich GCs are associated with the {\it in-situ} phase of galaxy formation, and that metal-poor GCs are all accreted. 
However, the situation for the MR and MP subpopulations may be more complex. Here, colour gradients in the subpopulations offer some clues. Radial colour gradients, in the sense that the innermost GCs are the most metal-rich,  have been detected in the MR and MP subpopulations separately in over a dozen GC systems (e.g. Harris 2009). They are relatively weak and hence require high quality data for a large sample of GCs to be detectable (which generally restricts galaxies beyond the Local Group to be massive ellipticals). These colour gradients correspond to metallicity gradients with average values of around --0.20 dex per dex. Within a given galaxy, the MR and MP subpopulations tend to have similar gradients (Forbes et al. 2011). There are hints that these gradients  correlate with host galaxy mass (Pastorello et al. 2015). At very large radii, the  gradients appear to flatten out to a constant colour/metallicity (which in turn can be used to infer the typical mass of accreted galaxies;   e.g. Arnold et al. 2011).


If the inner GC metallicity gradients of both subpopulations are associated with the {\it in-situ} phase of galaxy formation; then that process needs to produce GC subpopulations that have different spatial distributions and mean metallicities.  
An alternative possibility is that both MR and MP GCs were deposited into the inner regions by progressively more massive satellites (and their more metal-rich GCs), thus building up metallicity gradients in both GC subpopulations. A combination of this accretion process and {\it in-situ}  formed GCs is also possible.  For GCs in the outer halo regions, which show no radial colour gradients, a purely ex-situ/accretion origin seems most likely.   It will be interesting to see if  future simulations that incorporate GCs can reproduce  these behaviours in the blue and  red GC colour gradients, and hence provide a deeper insight into the processes and locations of GC formation.  It will also be useful to have more studies of GC systems and their host galaxies to large radii,  looking for common features in their GC/field population colour gradients and surface density/brightness profiles.  

Local Group galaxies provide a unique and complementary view of {\it in-situ} and accreted GCs, although one should bear in mind that it is a single environment, with a limited mix of galaxy types and masses. For example, the Local Group does not contain any massive elliptical galaxies, typically found in denser environments, which are thought to 
have assembled a large fraction of their mass by accreting satellite galaxies (and their GCs). 
Within the Local Group, GCs either fully or partially resolve into stars -- they can thus be readily identified on the basis of morphology alone, permitting much cleaner and more complete samples of GCs at large radii, and one can derive precise information about the constituent
stars from their CMDs. Studies to date have provided a wealth of evidence that accretion has played a significant role in building the halo GC systems of both the Milky Way and M31.

Searle \& Zinn (1978) were the first to argue that some fraction of the Milky Way GC system has an external origin. They showed that GCs outside
the solar circle exhibit no metallicity gradient, nor any correlation between horizontal branch (HB) morphology (taken as a proxy for age)
and metallicity, and concluded that the outer halo GC system arose due to the merger and accretion of `protogalactic fragments' in a slow
chaotic fashion.  Modern studies with much improved datasets have only strengthened this assertion, and indeed have placed it within the
cosmological context of hierarchical structure formation.

Arguably, the best example of GC accretion within the Milky Way is provided by the Sgr dwarf galaxy.  There is direct evidence that at least 5 GCs have been brought into the halo as part of this disrupting system, with several other promising candidates (e.g. Bellazzini et al. 2003; Law \& Majewski 2010; Massari et al. 2017). While most of these are old, metal-poor GCs, they also include some relatively young and metal-rich objects such as Terzan 7 ([Fe/H] = --0.56) and Whiting 1 ([Fe/H] = --0.65). Thus the Sgr dwarf is contributing {\it both} metal-poor and metal-rich GCs to the Milky Way halo, likely reflecting the age-metallicity relationship of the dwarf galaxy itself (Law \& Majewski 2010; Forbes \& Bridges 2010).  Less direct but still compelling evidence for GC accretion in the Milky Way comes from considering the distribution of metallicities, velocities, ages, HB morphologies and sizes of halo GCs (e.g. Mackey \& Gilmore 2004; Mar\'in-Franch et al. 2009; Forbes \& Bridges 2010; Dotter et al. 2010).  Unfortunately, the  number of
accreted GCs within the Milky Way today is highly uncertain, with estimates ranging from $\sim30-100$ (e.g. van den Bergh 2000; Mackey \& Gilmore
2004; Forbes \& Bridges 2010; Leaman et al. 2013).  Nonetheless, considering the total population of Milky Way GCs is $\sim 160$, these
numbers already demonstrate that a substantial fraction (up to 2/3) of the Milky Way's system is likely of an external origin.


In M31, the INT/WFC Survey, the SDSS and the Pan-Andromeda Archaeological Survey (PAndAS) have led to the discovery of more than
90 GCs in the outer halo of M31, lying at projected radii $>25$~kpc (e.g. Huxor et al. 2008, 2014; di Tullio Zinn \& Zinn 2013).  It is fascinating to note that this represents roughly 7 times as many GCs as in the comparable region of the Milky Way halo, and that many of the M31 halo GCs exhibit a striking correlation with faint stellar debris streams -- $`$smoking gun' evidence that they have been accreted along with their now-disrupted host galaxies.  Alignments between GCs and stellar streams are seen in both position and velocity space (Mackey et al. 2010, 2014; Veljanoski et al. 2014) and analysis suggests that $>60$\% of the outer halo M31 GC system has an accretion origin. Taken as a whole, the M31 halo GCs are remarkably uniform in colour and high resolution spectroscopic follow-up of a small sample has indicated that most of these are metal-poor, with [Fe/H]$\lesssim-1.5$ (Colucci et al. 2014; Sakari et al. 2015).  Nonetheless, accreted GCs can often be distinguished on the basis of their rather red HBs, which when combined with their metallicities, can only be explained by having younger ages (e.g. Mackey et al. 2013). Indeed, in this sense
the accreted members of the M31 GC system share similar characteristics to those in the halo of the Milky Way.

Recently, there have been concerted efforts to expand our knowledge of GCs in dwarf galaxies as these represent the most likely
donors of the accreted GCs in large galaxy halos.  The Local Group has more than 80 dwarf galaxies (M$_V>-16$) but only a handful of
these are currently-known to possess GCs.  Dedicated searches have improved our census in several cases (e.g. Hwang et al. 2011; Veljanoski et
al. 2013) while some dwarf galaxy GCs have been stumbled upon serendipitously (e.g. Cusano et al. 2016; Cole et al. 2017).  
Recently GCs have been found around very low mass dwarf satellite galaxies of the Milky Way and M31, i.e.  Eridanus II (M$_V=-7.1$; Crnojevic et al. 2016) and And I (M$_V=-11.7$; Caldwell et al. 2017).
There is clearly much to be learned from detailed study of GCs in the lowest mass galaxies (e.g. Zaritsky et al. 2016; Amorisco 2017;  Contenta et al. 2017), and from quantitative comparison of these objects with the suspected accreted  GC populations in the Milky Way and M31 halos.

\section{Are ancient GCs different from more recently formed GCs?}
Given the clear mass and density differences between Galactic open and globular clusters, early work on the origin of star clusters invoked special conditions of the early Universe for the formation of globular clusters.  The first detailed attempt to explain GC formation within a cosmological context was that of Peebles \& Dicke (1968), who noted that the typical masses of GCs are comparable to the Jeans mass shortly after recombination. Fall \& Rees (1985) later suggested that GCs might form as a result of thermal instabilities in collapsing protogalaxies. In both of these scenarios, GC formation was thus viewed as a special phenomenon of the early Universe, different from present-day star formation. The discovery of young ``super star clusters'' in the local Universe with masses and density equal to or even far above GCs (e.g. Holtzman et al.~1992), shifted the focus to scenarios in which GCs form largely by normal star formation processes.


%
%

In recent years, the differences between stellar populations in GCs and lower-mass open clusters, and in many cases, young/intermediate age clusters in the LMC/SMC, have been used as an argument that GCs may be unique after all.  In particular, GCs host star-to-star abundance spreads in C, N, O, Al, and He (a.k.a. multiple populations) whereas younger systems appear to be largely consist of a single population with homogeneous chemistry (see Bastian \& Lardo 2018). It should be noted that the phenomenon of multiple populations is present in both extremely metal-poor  (e.g. NGC~7078, NGC~7099; Carretta et al.\ 2010) and metal-rich GCs (e.g. 47 Tuc, NGC~6388).  Recent work, however, has shown that multiple populations are in fact present in GCs down to an age of $\sim2$~Gyr, $z_{\rm form} \approx 0.17$ (Hollyhead et al.~2017; Niederhofer et al.~2017; Martocchia et al. 2017).  While multiple populations have not been found below this age, it is doubtful that the GC formation process was significantly different before/after the $z_{\rm form} \approx 0.17$ boundary. The cause of this $\sim2$~Gyr age threshold in the onset of multiple populations is still not clear. It is possible that the phenomenon only manifests itself below a certain stellar mass (which translates into a certain age if only the RGB is investigated).  This could be due to the presence of only low-mass stars with abundance anomalies or possibly there is a so-far-unknown stellar evolutionary effect. For instance, the stellar mass boundary separating GCs with/without multiple populations straddles the boundary where stars have strong magnetic fields and also very different rotational properties. 

Instead of adopting special conditions for GC formation, which would imply a difference between them and younger star clusters with very similar properties (mass, size, density, stellar populations), a simpler approach would be to look for a global model of massive cluster formation. Can such a model explain the plethora of GC population properties, their relation to their host galaxy and the propeties of young/intermediate massive clusters?  There has been work in this direction, with some success (e.g. Kruijssen~2015).  Additionally, there is ongoing work to include what is known about massive cluster formation, evolution and destruction based largely on young massive clusters in the local Universe, and in cosmological hydrodynamical simulations of galaxy assembly (e.g.  Li et al. 2017; Renaud et al. 2017; Pfeffer et al.~2018)

As mentioned earlier, recent studies using gravitational lensing have been able to resolve down to nearly GC scales of 20-40~pc (Vanzella et al.~2017; Johnson,~T.~L. et al.~2017) in galaxies at redshifts $z=2-6$, offering the possibility of seeing GC formation in action.  While these studies are still preliminary, the properties of the observed star clusters 
are similar to those of Young Massive Clusters (YMCs) observed in nearby (i.e. low redshift) starburst galaxies in terms of their (unresolved) sizes, masses and relation to the star formation rate of their host galaxies. This suggests that, at least on the high-mass end, YMCs and proto-GCs may share similar formation mechanisms/properties. However, the detailed stellar population properties (e.g. presence of multiple populations or not) of these high-$z$ star clusters will likely remain unknown for the foreseeable future.

\section{What was the initial mass of a globular cluster system?}
\label{sec:icmf}

\subsection{Stellar population considerations}
The total integrated magnitude of the (halo) globular clusters in the Milky Way (here defined as those with $\mathrm{[Fe/H]}<-1$) is about $M_V(\mathrm{tot}) = -12.9$ (using data from Harris 1996). For a typical mass-to-light ratio of $\Upsilon_V\approx1.5$ (McLaughlin 2000; Strader et al. 2011), this corresponds to a total mass of about $1.2\times10^7 M_\odot$. The total mass of the stellar halo is estimated to be $(4-7)\times 10^8 M_\odot$ for $1 < R < 40$ kpc (Bland-Hawthorn \& Gerhard 2016), so the Milky Way GC system currently accounts for about 2--3\% of the stellar halo mass. However, because of dynamical evolution, GCs are expected to lose mass over time, and some clusters may have completely disrupted. It is thus likely that the initial mass of the GC population was significantly higher than the present-day observed value. 


The mass functions of young star cluster populations are typically well described by power-laws with a slope of $\approx -2$ at the low-mass end, and an exponential cut-off at the high-mass end, $\mathrm{d} N/\mathrm{d} M \propto M^{-2} \exp(-M/M_c)$ (see the review of Portegies Zwart et al. 2010). This is in contrast to the mass functions observed in old GC systems, which have a deficit of low-mass clusters with $\mathrm{d} N/\mathrm{d} M \approx$ const below some characteristic mass. When plotted in bins of $\log(M)$, or absolute magnitude, a flat ${\rm d}N/{\rm d}M$ gives rise to the familiar peaked form of the GC luminosity function with a peak at $M_V\approx-7.5$. The mass functions of old GCs can generally be quite well described by an \emph{evolved} Schechter functions of the form
\begin{equation}
\frac{\mathrm{d} N}{\mathrm{d} M} \propto (M+\Delta M)^{-2} \exp\left(-\frac{M + \Delta M}{M_c} \right) 
\label{eq:eschect}
\end{equation}
(Jord{\'a}n et al. 2007). Here, $M_c$ is again the high-mass cut-off and $\Delta M$ is the average amount of mass lost from each GC. Clusters with (initial) masses less than $\Delta M$ have been fully disrupted.
This simple formula describes the mass function of surviving GCs that started from an initial Schechter mass function (i.e. $\Delta M=0$), under the assumption that the mass-loss rate is independent of cluster mass (this assumption is unlikely to be perfectly valid). For the Milky Way, a fit to the GC mass distribution yields $\Delta M = 2.5\times10^5 M_\odot$ and $M_c = 8\times10^5 M_\odot$. 

If we assume that each of the \emph{surviving} $\approx$100 halo GCs in the Milky Way has lost 250,000 $M_\odot$, this amounts to about $2.5\times10^7 M_\odot$. Within GCs, the enriched fraction is usually about 50\%, so we may expect about $1.3\times10^7 M_\odot$ of the stars that have been lost from surviving GCs to have enriched composition.
It is interesting to compare this number with the amount of halo stars that display chemical abundance patterns typical of ``enriched'' stars in GCs. Martell et al. (2011) estimate that about 3\% of the stars in their sample of 561 halo giants display enhanced N and depleted C abundances.  If this fraction is representative of the halo as a whole, there should be about $(1.2-2.1)\times10^7 M_\odot$ of enriched stars in the halo.  Considering the "back-of-envelope" nature of this calculation, the similarity of these numbers of striking, and suggests that most of the chemically anomalous stars now observed in the halo field might have been lost from extant GCs. The above line of reasoning led Kruijssen (2015) to conclude that the minimum cluster mass needed for survival over a Hubble time is similar to the minimum mass for hosting multiple populations. Whether this is a coincidence or emerges from similar physics is an open question. In reality, the enriched fraction in the stars lost from GCs may be lower than  the present-day observed fraction within GCs, since the enriched stars tend to have a more centrally concentrated radial distribution. On the other hand, it would not be surprising if some fully disrupted clusters have contributed enriched stars to the halo.

The total current ($M_\mathrm{cur}$) and initial ($M_\mathrm{init}$) mass of a GC system can be found by integrating Eq.~(\ref{eq:eschect}) over all masses, setting $\Delta M=0$ for the initial mass. For a lower mass limit of 100 (5000) $M_\odot$, the ratio $M_\mathrm{init}/M_\mathrm{cur}$ is 24 (13). Note that this does not include the factor $\sim2$ that each GC will lose due to stellar evolution. For more realistic assumptions about the dynamical evolution, Kruijssen \& Portegies Zwart (2009) found factors of $M_\mathrm{init}/M_\mathrm{cur}$ = 64 (39), including the effects of stellar evolution. Several other authors have found $M_\mathrm{init}/M_\mathrm{cur}>10$ (Fall \& Zhang 2001; Vesperini 2001; Jord{\'a}n et al. 2007), implying that disrupted GCs might account for a significant fraction of the stars now belonging to the halo field. This provides an indirect constraint on the presence of multiple populations in the fully disrupted GCs: if they contained chemically anomalous stars in the same proportion as surviving GCs, these stars should now be present in the field in far greater numbers than are actually observed. Hence, it appears likely that if the present-day GC mass function is the result of dynamical evolution from an initial Schechter-like function, the fraction of chemically anomalous stars in the disrupted (primarily low-mass) GCs was lower than in present-day surviving GCs. This would be in accordance with the tendency for the enriched fraction to increase with GC mass (Milone et al. 2017), and the fact that multiple populations have not been found in open clusters.
%
%

Of course, it may be premature to assume that GC systems formed with a Schechter-like mass function. In particular, the large amount of mass loss required to turn a Schechter function into the present-day MF is in tension with the large fractions of metal-poor stars that \emph{presently} belong to GCs in dwarf galaxies like the Fornax dSph. At metallicities $\mathrm{[Fe/H]}<-2$, 20-25\% of the stars in the Fornax dwarf belong to GCs (Larsen et al. 2012), and similar high fractions are found in the WLM and IKN dwarfs (Larsen et al. 2014).
This appears difficult to reconcile with the idea that the present-day GCs should represent less than 10\% of the initial mass of a GC population that initially followed a Schechter mass function. Alternatively, the initial mass distribution of GCs may have been more top-heavy, for example by shifting the lower mass cut-off to higher masses at the time of GC formation and/or invoking a shallower slope. Indeed, the young star cluster populations in some star-forming dwarf galaxies are dominated by one or two very massive clusters (e.g. NGC~1569, NGC~1705, NGC~1023-A).

Of course, the high GC/field ratios in dwarf galaxies also pose a severe challenge for scenarios that invoke large amounts (90--95\%) of mass loss from individual clusters in order to explain the large fraction of chemically anomalous stars within the GCs (D'Ercole et al. 2008; Schaerer \& Charbonnel 2011; Bekki 2011; Cabrera-Ziri et al. 2015). This challenge is greatly compounded if one additionally has to account for the dissolution of lower-mass GCs. 
It would be interesting to determine the fraction of enriched field stars in dwarf galaxies such as Fornax. If the GCs there have experienced significant mass loss, we might expect the enriched fraction to be much higher in the dwarfs than in the Milky Way halo. 
\subsection{The globular cluster mass function: nature or nurture?}
\newcommand{\mto}{M_{\rm TO}}
\newcommand{\msun}{{\rm M}_\odot}
\newcommand{\mnras}{\textit{MNRAS}}
\newcommand{\apj}{\textit{ApJ}}
\newcommand{\ndotrel}{\dot{N}_{\rm rel}}
\newcommand{\mdotev}{\dot{M}_{\rm ev}}
\newcommand{\rg}{R_{\rm G}}

\newcommand{\kpc}{\rm kpc}
\newcommand{\kms}{\rm km/s}
\newcommand{\rhoh}{\rho_{\rm h}}
\newcommand{\myr}{\rm Myr}
\newcommand{\vcirc}{V_{\rm circ}}

The evolved globular cluster mass function (GCMF) mentioned in equation~(\ref{eq:eschect}) does a good job in describing the luminosity function of Milky Way GCs, and those of GCs in external galaxies. But we have yet to  address the disruption mechanism that is responsible for shaping the GCMF. An important question is whether it is able to produce the correct turn-over mass ($\mto$), i.e. the value of $\Delta M$, roughly independent of environment, as observed. Jord\'{a}n et al. (2007), Villegas et al. (2010), and Harris et al. (2014) discuss subtle dependencies of the shape of the GCMF on galaxy mass. Carretta et al. (2010) show that the luminosity function of outer halo clusters contains slightly more faint (i.e. low mass) GCs than the inner halo, and the luminosity function is similar to that of GCs in Local Group dwarf spheroidals. This higher frequency of low-mass GCs may point at evolution in weaker tidal environments.

Baumgardt \& Makino (2003) presented a suite of $N$-body simulations of GCs dissolving in a Galactic tidal field, which are still the default reference for GC evolution. The models consider GCs with  a  stellar mass spectrum, the effects of stellar evolution and  orbits with various eccentricities in a Galactic potential that is approximated by a singular isothermal halo (i.e. $\rho(\rg)\propto \rg^{-2}$). They find thatthe mass-loss rate as the result of two-body relaxation driven evaporation in a tidal field, $\mdotev$, depends on $M$ and the tidal field as $|\mdotev|\propto M^{x}\vcirc/\rg$, where $\vcirc$ is the circular velocity in the galaxy and  $1/4\lesssim x\lesssim1/3$. 
Using their results for the most massive GCs in their suite ($\sim10^5\,\msun$), and ignoring the small $M$ dependence of $\mdotev$ and adopting $\vcirc\simeq220\,\kms$, their results imply $\Delta M(\rg) \simeq 2.4\times10^5\,\msun\,(\kpc/\rg)$ at an age of 12 Gyr. This means that at $\rg\simeq 1\,\kpc$ the GCMF could have evolved from a $-2$ power-law function to a peaked function as the result of evaporation, but these models predict a decreasing $\mto$ with increasing $\rg$. In the power-law regime of the GCMF, $\mto\propto\Delta M$ (Jord\'{a}n et al. 2007), hence at $\rg\simeq10\,\kpc$, $\mto$ would be a factor of 10 too low. Including the $M^{x}$ dependence makes the discrepancy at large $\rg$ even larger (Gieles 2009). A  concern with this argument is that beyond $\rg\gtrsim10\,\kpc$, the majority of the GCs are likely accreted from disrupted dwarf galaxies (see \S~\ref{sec:formlocation}). We therefore need to understand the dynamical evolution of GCs within the framework of hierarchical galaxy formation to address the GCMF problem. 

Typical values for the tidal field in dwarf galaxies are $\vcirc\simeq10\,\kms$ and $\rg\simeq1\,\kpc$, such that $\vcirc/\rg \simeq 10\,\myr^{-1}$, i.e. comparable to the Milky Way tidal field at $20\,\kpc$, i.e. a tidal field that is too weak to get the correct $\mto$ by evaporation.  In addition, the (dark matter) density profile in dwarfs is flatter than $-2$, which further reduces $\mdotev$  by a factor of $2$ or $3$ compared to a singular isothermal halo (Claydon et al. 2017). Finally, the adiabatic growth of the Milky Way brings GCs closer to the Galactic centre, such that $\mdotev$ as derived from their present day orbit overestimates the  $\mdotev$ averaged over their evolution in the Milky Way (Renaud \& Gieles 2015). Therefore, the GCMF problem gets worse when invoking the hierarchical growth of galaxies.

Several models have tried to explain the universality of the GCMF by internal processes, such as the expulsion of residual gas (Baumgardt et al. 2008) or stellar evolution (Vesperini \& Zepf 2003), but these models make predictions for YMCs that have not been observed (see the review of Portegies Zwart et al. 2010). McLaughlin \& Fall (2008) propose that evaporation results in mass loss on a relaxation time-scale, such that $|\mdotev|\propto \rhoh^{1/2}$, where $\rhoh$ is the density within the half-mass radius. In this model the details of the tidal field do not enter and this naturally results in an $\rg$-independent $\mto$, and a correlation between $\mto$ and $\rhoh$, which is indeed observed. Although this model successfully explains the observations, the required mass-loss dependence on cluster properties -- independent of environment --  is inconsistent with results of numerical simulations of cluster evolution (e.g. Lee \& Ostriker 1987; Gieles \& Baumgardt 2008).  
Gieles et al. (2011) show that a more plausible origin for the correlation between $\rhoh$ and $M$ is relaxation driven expansion of low-mass GCs, rather than preferential disruption of high-density clusters. 

Elmegreen (2010) and Kruijssen (2015) propose that interactions with giant molecular clouds (GMCs) in the first few 100 Myrs to Gyrs could efficiently disrupt low-mass GCs and shape the GCMF by dynamical evolution, but soon after GC formation. These `tidal shocks' with GMCs disrupt the cluster on  a time-scale that is proportional to the density of the cluster (Spitzer 1958). This only results in mass dependent disruption, if the density of clusters correlates with their mass (Gieles et al. 2006). This appears indeed to be the case for YMCs (Larsen 2004), which makes interactions with GMCs an additional  disruption mechanism that can `turn over' the GCMF. Elmegreen (2010) and Kruijssen (2015) adopt a weak mass-radius correlation (or a constant radius) and use estimates for the ISM properties at high redshift to argue that the strength of shock disruption is sufficient to get the correct $\mto$ within $1$--$3\,$Gyr. Gieles \& Renaud (2016) showed that the removal of mass by tidal shocks results in an increase of the cluster density, i.e. the disruption by tidal shocks is a self-limiting process. 
When considering the combined effects of tidal shocks (which shrinks clusters) and relaxation (which expands clusters), an equilibrium mass evolution $|\dot{M}| \propto M^{1/3}$ is found (Gieles \& Renaud 2016), similar to the mass evolution due to relaxation in a steady tidal field. 
We note that Kruijssen~(2015) already considered more compact clusters when studying the evolution of the mass function, in a way that is consistent with the decrease of cluster radii due to their response to tidal shocks from Gieles \& Renaud~(2016), finding that tidal shocks are still able to turn over the GCMF. 
Whether the ISM  at high redshift is sufficiently dense to evolve the GCMF with tidal shocks across the entire galaxy mass range, is an open question.

As some level of fine-tuning is required to obtain a near universal GCMF with dynamical evolution it is tempting to consider an explanation that relies more on nature, rather than nurture. Indeed, several older models for the formation of GCs suggested  that a typical mass of $\sim10^6\,\msun$ exists  (Peebles \& Dicke 1968; Fall \& Rees 1985; Bromm \& Clarke 2002). These were often based on cooling instability and Jeans mass arguments, and required fragmentation to produce lower mass GCs. More recent works
(e.g. Glover \& Clark 2014), 
with updated cooling physics,  suggest that the situation is 
more complicated than initially thought. 
We note that the majority of GCs in the Milky Way belong to the metal-poor category, thus a formation model  involving inefficient cooling at low metallicity is attractive.
However, it is not clear if the observed floor in GC metallicity,  i.e. no GCs have been observed with [Fe/H] $<$ -2.5, can be reproduced by cooling instability physics. Molecular and atomic line cooling, which are dominant in the star-forming ISM, do not differ fundamentally from local-Universe cooling until much lower metallicities (e.g. Glover \& Clark 2014).

A way forward to address the long standing problem of the GCMF is to expand the modelling efforts to include the radius/density of the clusters. The disruption by tidal shocks are efficient in destroying clusters of low density, whereas $\mdotev$ depends only on $M$ and the stationary tidal field. Fast cluster evolution models, such as \texttt{EMACSS} (Alexander \& Gieles 2012), guided by cosmological zoom-in simulations are ideally suited for enabling globular cluster population synthesis studies that aim to reproduce the density of GCs in the mass, density, Galactocentric radius and [Fe/H] space.

\section{How did globular clusters form?}

Broadly speaking, the current literature contains two families of models for the formation and evolution of GC systems in the context of galaxy formation. The first family of models associates GC formation with special conditions in low-mass dark matter haloes, during or before reionisation (e.g.~Peebles \& Dicke 1968; Katz \& Ricotti 2014; Trenti et al.~2015; Kimm et al.~2016). The second family of models considers GC formation a natural byproduct of the active star formation process (which is often linked to high gas pressures) seen at high redshift (Elmegreen 2010; Shapiro et al. 2010; Kruijssen 2015). In this latter branch of models, there is an ongoing debate between studies finding a relatively high importance of galaxy mergers in producing GCs (Li et al.~2017; Renaud et al.~2017; Kim et al.~2017) and those finding most GC formation proceeds in galaxy discs (Kruijssen 2015; Pfeffer et al.~2018).

An end-to-end description of the assembly of GC systems during galaxy formation requires a model for star cluster formation in the early Universe, their initial mass function, their dynamical destruction (both in high-redshift environments by tidal perturbations from dense gas clouds and in low-redshift environments by evaporation), and their redistribution during hierarchical galaxy assembly and growth. The aforementioned two families of GC formation models differ in almost all of these ingredients and sometimes fundamentally so. Therefore, we briefly review each of these elements in the context of both model families, before discussing the different numerical approaches that are used by recent models, and discussing the key model predictions.

\subsection{Globular cluster formation at very high redshift and connection to dark-matter halo assembly}

Globular clusters host some of the oldest stellar populations in our Galaxy and have been used in the past to constrain the age of the Universe (e.g. see Krauss \& Chaboyer 2003). Therefore, it is not surprising to see various investigations of formation scenarios at very high redshift, and connections between GC formation and dark-matter (DM) halo assembly. Early work by Peebles \& Dicke (1968) noted that the Jeans mass at high redshift ($\sim 10^6\,\mathrm{M_{\odot}}$) is comparable to the GC stellar mass, while other more recent studies focused on formation within DM mini-halos with characteristic mass $M_{DM}\sim 10^8\,\mathrm{M_{\odot}}$. For example, Padoan et al. (1997) suggested that compact star clusters can be formed through efficient H$_2$ cooling; Cen (2001) argued in favor of formation triggered by shocks induced by a hydrogen ionization front; Naoz \& Narayan (2014) investigated collapse of gas displaced by stream velocity from its parent DM halos; Trenti et al (2015) proposed instead that mergers of gas-rich but star-free mini-halos ($M_{DM}\sim 10^7\,\mathrm{M_{\odot}}$) can lead to shock-induced compression and formation of central stellar clusters that would later be stripped of their DM envelope and become the population of old galactic GCs with a wide spread in metallicity but narrow difference in age. The appealing features of these models are that they are based on fundamental physical processes that are well established to happen during the early history of the Universe, at a time when the conditions for star formation were markedly different compared to low-redshift star formation because (1) the characteristic mass for collapsed structure was significantly lower, hence the massive disk galaxies that are frequently hosting young star clusters at lower redshift were extremely rare;
(2) the merger rate was significantly higher than that of the local Universe (e.g., see Fakhouri et al. 2010) since density scales with (1 + z)$^3$; 
and (3) the ionizing background was lower (especially pre-reionization), but at the same time the higher cosmic microwave background floor was limiting the cooling of the gas to $T\gtrsim 30$ K, potentially affecting characteristic mass and fragmentation of gas clouds (e.g. Smith, Sigurdsson \& Abel 2008). However, all proposed ideas are relatively qualitative, and further work is needed to develop them to the point where a data-model comparison is effective in falsifying them. 

In addition, the typical mechanisms that are proposed for formation of compact stellar systems are not likely to be in place at lower redshift, thus these classes of models explicitly (or implicitly) assume that younger GCs form through a different channel, even though recent observations show that young GCs may have properties that are indistinguishable from old GCs (Hollyhead et al. 2016; Niederhofer et al. 2017). The existence of multiple channels could be physically motivated, since compact star clusters are forming in different contexts (e.g. nuclear star clusters, see Seth et al. 2006), but a single formation channel capable of explaining the broad range of GC system properties would arguably be preferred by Occam's razor. More broadly, current age measurements for Milky Way GCs are consistent with a significant fraction of GCs being formed during the first 800 Myr after the Big Bang (Section \ref{sec:when_gcs_form}), but systematic uncertainties dominate the error budget, preventing a strong inference on observed ages. The key development that would allow for an efficient discrimination between models would be a technique to measure ages with uncertainties below $500$ Myr to unequivocally establish what fraction of GCs (if any) are formed before the epoch of reionization at $z>6$. 

A second feature broadly shared by models of GC formation in the context of DM halo assembly is that a characteristic mass scale for the formation process is identified. Therefore, the broad predictions/expectations are that the initial cluster mass function has a preferred scale, e.g in the form of a log-normal distribution, rather than a power-law (e.g. see Trenti et al. 2015).

Finally, while some models argue for purely baryonic formation processes (e.g. Naoz \& Narayan 2014), the general feature of this class is that the dynamical assembly of the parent galaxy halo leads to tidal stripping of the DM envelope around GCs, which are expected to be characterized by high stellar densities ($\rho_c\sim 10^6~\mathrm{M_{\odot} pc^{-3}}$) and compact radii ($r_h\sim2$ pc) if formed at $z\sim 10$ (Devecchi \& Volonteri \ 2009; Ramirez-Ruiz et al.\ 2015). Under these conditions it would be realistic to expect that even when the process is efficient, some GCs would still survive embedded in part of their DM sub-halo. Thus, high-resolution simulations that follow the complex gravitational dynamics of repeated mergers and stripping events would be needed to progress quantitatively to testable predictions.

\subsection{Globular cluster formation as the natural outcome of star formation in normal high-redshift galaxies}

In the local Universe, YMCs with masses and radii (and hence densities) very similar to GCs are observed to form in gas-rich environments of high star formation rate (SFR) and gas pressure (for recent reviews, see Portegies Zwart et al.~2010; Longmore et al.~2014; Kruijssen 2014). Given that the gas fraction, gas pressure, SFR density, and SFR per unit gas mass in galaxies are lower at $z=0$ than in the early Universe, with the gas pressure and SFR peaking at $z \approx 1-3$ (e.g.~Swinbank et al.~2011, 2012; Madau \& Dickinson 2014; Schinnerer et al.~2016; Scoville et al.~2016; Tacconi et al.~2017), it is therefore sensible to consider that GCs are the products of regular star and cluster formation in normal high-redshift galaxies. Indeed, this idea has motivated a broad range of models in which GCs form in gas-rich, high-redshift galaxy discs (e.g.~Kravtsov \& Gnedin 2005; Elmegreen 2010; Kruijssen 2015). In some variants, the formation of young GCs is enhanced by galaxy mergers (e.g.~Li et al. 2017; Kim et al.~2017). The simulations do not extend yet to $z=0$ and therefore cannot predict whether the clusters formed in mergers survive until the present day (which at least for intermediate-mass clusters has been questioned in recent theoretical work, see Kruijssen et al.~2012).

From a theory perspective, the conditions in gas-rich galaxies at high redshift are favourable to GC formation for two reasons. Firstly, the high turbulent pressures are predicted to lead to a high fraction of star formation occurring in gravitationally bound clusters (Kruijssen 2012; also see Elmegreen 2008), which follows the same trends as observed in actively star-forming galaxies in the local Universe (Adamo et al.~2015; Johnson et al.~2016). Secondly, the high gas pressures in the turbulent interstellar medium (ISM) of these galaxies also imply high maximum mass-scales for gravitational collapse (e.g.~Agertz et al. 2009, Dekel et al.~2009), which potentially enables the formation of massive GMCs if they overcome shear, centrifugal forces, and feedback (e.g.~Reina-Campos \& Kruijssen 2017). Some fraction of these GMCs may form massive clusters (e.g.~Harris \& Pudritz 1994). In the optimal case, the resulting maximum cluster mass scales linearly with the GMC mass, modulo factors corresponding to the star formation efficiency and the fraction of star formation that ends up gravitationally bound (Kruijssen 2014). Observations and theory show that the resulting maximum cluster mass increases with the gas pressure (which may be traced by the gas or SFR surface density, Johnson,~L.~C. et al.~2017; Reina-Campos \& Kruijssen 2017). In the highly turbulent and high-pressure environment at high redshift, this plausibly leads to the formation of very massive stellar clusters that can remain gravitationally bound over a Hubble time.

It is not yet clear what the role of galaxy mergers is in determining the properties of the GC population at $z=0$. Galaxy merger rates are orders of magnitude higher at high redshift ($z>1$) than at the present day (e.g.~Fakhouri et al.~2010) and could facilitate the assembly of very large GMCs (either by direct cloud collisions or by raising the ISM pressure), which could host more massive star clusters than in discs (e.g.~Li et al. 2017; Kim et al.~2017). While the conditions in galaxy discs may already be sufficient to form GCs, mergers could therefore enhance GC formation. However, mergers might also increase the destruction rate of star clusters (Kruijssen et al.~2012) due to the increased gas densities (see below). In addition, the fraction of all star formation occurring in mergers may be low, of the order 10\% (Rodighiero et al.~2011). In view of these uncertainties, a detailed quantification of the contributions of mergers to GC formation is needed. Even if mergers do not represent a large fraction of all GC formation, they may eject GCs into galaxy halos where they can survive over longer timescales (Kravtsov \& Gnedin 2005; Kruijssen 2015).

It is now widely accepted that regular star cluster formation in galaxy discs leads to a power law initial cluster mass function with a gradual (e.g.,~exponential) truncation at the high-mass end; pure power laws are ruled out (Larsen 2009; Gieles 2009; Portegies Zwart et al.~2010; Li et al.~2017; Johnson,~L.~C. et al.~2017). Also, the GC mass function has a similar truncation at the high-mass end (Fall \& Zhang 2001; Jordan et al.~2007; Kruijssen 2015). Therefore, the favoured model for the initial cluster mass function during regular star formation in high-redshift galaxies is a Schechter (1976) type function. We note that log-normal initial cluster mass functions are inconsistent with the hierarchical ISM structure observed in galaxy discs (e.g.~Elmegreen \& Falgarone 1996; Guszejnov et al.~2017). In the specific context of GC formation as the natural outcome of high-redshift star formation that is considered here, the ICMF should thus follow a (truncated) power law. In a broader context, this is a general result of the scale-free structure of the ISM. However, the exact form of the GC initial cluster mass function remains unknown until it has been observed directly.

Even if GC-like objects form at high redshift, they need to survive till $z=0$ in order to be observed as GCs. Models and simulations accounting for this process need to include a variety of mass loss and destruction mechanisms, such as stellar evolution, tidally-limited evaporation, time-variable tidal perturbations (`tidal shocks') due to interactions with e.g.~the ISM or galactic structure, and dynamical friction. Stellar evolutionary mass loss equally affects star clusters of all masses, but evaporative and tidal shock-driven mass loss mostly affects low-mass clusters. Theoretical and numerical studies find that tidal shocks represent the dominant destruction mechanism in the gas-rich environments where star clusters form (Gieles et al.~2006; Lamers \& Gieles 2006; Elmegreen \& Hunter 2010; Kruijssen et al.~2011; Miholics et al.~2017). As a result, it is a key requirement for star cluster survival to escape its gas-rich formation environment, possibly by galaxy mergers or other (gas or cluster) migration mechanisms (Kravtsov \& Gnedin 2005; Kruijssen 2015). To date, few models for the origin of GCs in the context of galaxy formation have included tidal shocks, barring a couple of exceptions (Elmegreen 2010; Kruijssen 2015; Pfeffer et al.~2018). It is important to note that the efficiency of tidal shocks depends on the spatial resolution of the simulations. It always represents a lower limit unless the high-mass end of the GMC mass function is well-resolved, because a GMC mass function with a slope shallower than $-2$ implies that most of the destructive power should come from high-mass ISM structures.

By contrast to the other mass loss mechanisms, dynamical friction destroys the most massive star clusters. This has an important implication in the context of direct detections of young GCs with next-generation telescopes such as JWST (see Section~\ref{sec:when_gcs_form}). While the most massive GCs are likely to survive till $z=0$ against disruption by evaporation and tidal shocks, they are the least likely to survive inspiral by dynamical friction, provided that they formed in the inner parts of their host galaxy where most of the dense ISM resides. A $10^7\,{\rm M}_\odot$ cluster (which is just three times more massive than the present-day GC truncation mass) in a galaxy with a circular velocity of $100~{\rm km}~{\rm s}^{-1}$ has a dynamical friction timescale of $t_{\rm df}\sim5$~Gyr for an initial galactocentric radius of $R\sim5$~kpc, with $t_{\rm df}\propto R^2$, highlighting that sufficient time is plausibly available for this mechanism to affect the most massive GCs and to have set the maximum mass scale observed today. It is therefore unclear if the massive clusters observed with JWST will survive to become GCs at $z=0$; the identification of star clusters at high redshift as proto-GCs is challenging, because it requires accounting for the above disruption mechanisms.

When considering the balance between formation and destruction mechanisms, there have been ongoing debates whether the observables describing the GC population at $z=0$ are shaped mostly by formation or by disruption. Examples are the peaked GC mass function, which is an imprint of formation in GC formation models related to dark matter haloes and reionisation, whereas it is the result of GC disruption in the `regular cluster formation' family of models. Likewise, the specific frequency (number of GCs per unit galaxy mass or luminosity) has been referred to as a GC formation efficiency (e.g.~McLaughlin~1999; Peng et al.~2008), whereas recent work argued it mostly represents a survival fraction (Kruijssen 2015). The same question has been discussed in the context of the constant GC system mass per unit dark matter halo mass (e.g. Blakeslee et al.~1997; Blakeslee 1999; Spitler \& Forbes 2009), which has been suggested to be the result of self-regulation by feedback from the host galaxy (Hudson et al.~2014; Harris et al.~2015), GC formation proportional to halo mass (e.g.~Katz \& Ricotti 2014), or as a coincidental combination of the opposite galaxy mass dependences of the GC survival fraction and the baryon conversion efficiency, with possibly an additional linearising effect of the central limit theorem during hierarchical galaxy assembly (Kruijssen 2015). Finally, it is relatively broadly supported that the GC metallicity distribution largely traces formation, in the sense that MP GCs mostly formed ex-situ in low-mass galaxies, whereas MR GCs mostly formed in-situ within the main progenitor (Tonini 2013).

\section{Numerical approaches to model globular cluster formation during galaxy formation}

Attempts to describe GC formation mechanisms as an integral aspect of galaxy formation are rapidly increasing in number. Here, we briefly summarise the different modelling approaches in broad categories and provide a short description of their benefits and shortcomings.

\subsection{Extremely high-resolution cosmological zoom-in simulations}

As the ongoing development of computing algorithms enables galaxy formation simulations reach parsec-scale spatial resolution, it opens a new avenue to study GC formation by resolving the process directly. This requires particle or cell masses below $10^3\,{\rm M}_\odot$ and is computationally extremely demanding. As a result, it is presently very difficult to follow the system to $z=0$ and current examples in the literature therefore stop the simulation at a higher redshift. The major advantage of this approach is that the formation of clusters is based on a higher degree of `first principles' physics. Still, some important physical mechanisms remain uncaptured, most critically so in the area of star cluster mass loss, because the collisional stellar dynamics that drive evaporation and determine the cluster density profile (and hence its response to tidal shocks) are unresolved. It is also not possible to know if the star cluster may fragment into smaller systems when increasing the resolution. However, direct high-resolution simulations are the best avenue for testing and exploring the physics of massive star clusters in high-redshift environments.

Kim et al.~(2017) present the first cosmological hydrodynamic simulation of the formation of a massive star cluster ($M \sim 10^6\,M_\odot$) at redshift $z \approx 7$, resolved with many ($\sim10^3$) stellar particles. Similarly, Li et al.~(2017) model the formation of GMCs in cosmological simulations with about 5 pc spatial resolution, and the growth of cluster particles through gas accretion that is terminated by self-consistent feedback. This method enables the particle masses to be interpreted as cluster masses and test the simulation results against the observed cluster mass function. The simulation finds good agreement, with a power law initial cluster mass function and an exponential truncation at the high-mass end.

\subsection{Subgrid modelling in self-consistent galaxy formation simulations}

The second method for modelling GC formation and evolution during galaxy formation is to use subgrid models. The major advantage of this approach is its modest computational cost and, as a result, the greatly improved statistics of the modelled GC population, while retaining some description of the pertinent physics. These models typically adopt numerical resolutions similar to other galaxy formation simulations and, as a result, can comfortably run to $z=0$, enabling comparisons between the modelled and observed GC populations. In addition, the shorter run time enables carrying out large numbers of simulations that systematically explore the effects of the subgrid models describing cluster formation and evolution, providing insight into which physics matter most in shaping the GC population. This method also opens up the possibility of simulating a range of galaxies with a variety of formation and assembly histories, thus probing the impact of these histories on the properties of the GC population and thereby fulfilling the potential of GCs as tracers of galaxy formation and evolution. The relatively low computational cost also permits simulating volumes larger than Milky Way-mass galaxies, feasibly up to Virgo cluster-like masses with current computational facilities. The obvious downside of this approach is that subgrid models require compromises in terms of the accuracy and level of detail at which the formation and disruption physics are described. However, the fact that at least some physically-motivated description of the GC formation and disruption physics is included implies that this approach is currently the only method that provides an end-to-end description of the formation and evolution of the entire GC population during galaxy formation, from high-redshift till the present day, while capturing the environmental dependences and complete population statistics.

Initially, the modelling efforts in this direction post-processed dark matter-only simulations or the results thereof by adding baryons through observed scaling relations and analytic physical prescriptions and inserting GCs at particular epochs. These models are largely (semi-)analytical in nature (e.g.~Muratov \& Gnedin 2010; Li \& Gnedin 2014; Kruijssen 2015, Choksi et al.~2017). More recently, self-consistent hydrodynamic simulations of galaxy formation are used to model the formation and destruction of the GC population over long (from many Gyr to a Hubble time) timescales (Kruijssen et al.~2011, 2012; Pfeffer et al.~2018; Kruijssen et al.~2018). Both branches of approaches highlight the major impact of environment on the GC system properties, implying that accounting for the galaxy formation and assembly history is crucial for understanding GC formation and evolution. An important consequence of this variance in GC systems is that a single simulation of a single galaxy is unlikely to provide widely applicable insights -- statistical samples of galaxies with a range of assembly histories are needed to come to more generally applicable conclusions.

\subsection{Particle tagging and dark matter-only simulations}

The third approach by which the formation and assembly of GC systems is studied is by particle tagging in (possibly dark matter-only) galaxy formation simulations. Tagging particles and representing these as GCs is the most phenomenological approach out of the three methods described here, which is why it has been popular for more than a decade (see e.g.~Moore et al.~2006, but also more recent examples by Corbett Moran et al.~2014; Mistani et al.~2016; Renaud et al.~2017). Its most obvious advantage is the major versatility in selecting which subpopulation of particles to tag, and the fact that the tagging can be performed in post-processing, whereas the two other methods discussed above require on-the-fly modelling. The downside is that models relying on particle tagging do not include a self-consistent treatment of the relevant physics and will yield GC systems that are biased towards the tagged (e.g.~field star or dark matter) population. They also have trouble to self-consistently capture cluster disruption for all clusters across a large sample of galaxies due to the prohibitively large amount of storage space required.

The above limitations are particularly relevant because the environmental dependence of GC formation and disruption physics implies that the spatial distribution of GCs at each redshift matters for their properties at $z=0$. For instance, the specific frequency of GCs varies by up to three orders of magnitude across the GC metallicity range (Lamers et al.~2017), suggesting that, combined with the stellar metallicity gradient in galaxies, on average, field stars do not occupy the same distribution in six-dimensional (position-velocity) phase space as GCs. Therefore, the results of particle tagging methods will be sensitive to the exact subpopulation of particles that is tagged.
Likewise, if satellite dark matter haloes are tagged by following the observed spatial profiles of GCs (as in Moore et al.~2006), it suggests GC formation redshifts that are too high to be consistent with the observed GC ages (cf.~Brodie \& Strader 2006; Forbes \& Bridges 2010). Most likely, this discrepancy arises because multiple GCs could form per satellite halo and because the spatial profiles of GCs were affected by environmentally dependent GC formation efficiencies and survival rates.

Despite the various caveats that should be kept in mind when using particle tagging to model GC systems, this approach can lead to valuable insights when physically-motivated, specific subsets of particles are tagged. For instance, Tonini (2013) used dark matter-only simulations to demonstrate how bimodal metallicity distributions of GCs can arise due to the accretion of dwarf galaxy satellites. Renaud et al. (2017) showed that the old star particles ($>10\,$Gyr) in a cosmological zoom-in simulation have a similar density distribution and kinematics at low redshift as found for the Milky Way's oldest GCs. 
This suggests that GCs may follow the earliest phases of star formation, where some (yet to be understood) fraction of the stars end-up in globular clusters. Such studies should be followed up with detailed subgrid models incorporating GC formation and evolutionary processes.

\section{Prospects for modelling the formation and co-evolution of GCs and their host galaxies in a cosmological context}
\label{sec:prospects}

In order for cosmological hydrodynamical simulations of galaxy formation to prove an informative tool with which to study GC formation and evolution, an obvious prerequisite is that they should reproduce key observed scaling relations of the galaxy population. Only in recent years have such calculations begun to satisfy this condition, largely thanks to significant developments in the modelling of "feedback" processes (e.g. Larson 1974; White \& Rees 1978; Dekel \& Silk 1986) that regulate (and even quench) galaxy growth by (re)heating and ejecting cold gas from galaxies. 

Owing to the dynamic range of the problem, the scope of cosmological simulations is often limited by their memory footprint, and those that follow volumes sufficiently large to realise a representative population of galaxies typically achieve a spatial resolution of only $\sim 1$ kpc. This precludes detailed numerical modelling of the life cycle of the ISM, motivating instead phenomenological treatments of star formation (see e.g. Springel \& Hernquist 2003; Scannapieco et al. 2006; Schaye \& Dalla Vecchia 2008) and feedback (see e.g. Katz 1992; Navarro \& White 1993; Springel, Di Matteo \& Hernquist 2005, Stinson et al. 2006; Dalla Vecchia \& Schaye 2008;  Hopkins et al. 2012; Agertz et al. 2013; Booth et al. 2013; Keller et al. 2014). 

The use of such treatments has shown, however, that the inclusion of feedback does not necessarily lead to the formation of realistic galaxies; simulations incorporating well-motivated feedback prescriptions can yield galaxy populations with unrealistic masses, sizes and kinematics (e.g. Crain et al. 2009; Lackner et al., 2012). The systematic study of Scannapieco et al. (2012) found that leading hydrodynamical models, when applied to identical initial conditions of a halo much like that of the Milky Way, yielded present-day stellar masses that varied by almost an order of magnitude. It is likely the case that much of this dispersion stems from unintentional numerical losses of the injected feedback energy (for example in the regime where the local cooling time of a gas element is shorter than its sound crossing time, see e.g. Dalla Vecchia \& Schaye 2008), as well as differences in the assumed physical efficiency of the feedback. The dispersion serves to illustrate the sensitivity of the global star formation history of galaxies to the efficiency of feedback. Recognition of this sensitivity has motivated exploration of the impact of varying the properties of feedback-driven outflows in response to the growth of the galaxy, which has been found to be an effective means of governing star formation histories (e.g. Oppenheimer et al. 2010; Schaye et al. 2010; Puchwein \& Springel 2013). This has led to the development of simulations that explicitly calibrate feedback efficiencies to reproduce key low-redshift observables such as the galaxy stellar mass function from Illustris (Vogelsberger et al. 2014)  and from EAGLE simulations (Schaye et al. 2015; Crain et al. 2015). Encouragingly, these simulations also reproduce many other diagnostic quantities that were not calibrated.

Realistic galaxy formation simulations present a promising foundation from which to explore the formation and co-evolution of GCs and their host galaxies. Renaud et al. (2017) examined a high resolution cosmological simulation of a galaxy evolved to $z=0.5$. Tagging stellar particles as potential star clusters, they computed the tidal forces these particles were subjected to throughout the assembly of the host galaxy. They showed that the tidal forces experienced by stellar particles evolve markedly over cosmic timescales, and as a function of a particle's (evolving) location within the galaxy and its progenitors. The E-MOSAICS project (Pfeffer et al.~2018; Kruijssen et al.~2018) couples the analytic MOSAICS models (Kruijssen et al.~2011, 2012) of star cluster formation and evolution to the EAGLE galaxy formation model, initially presenting results from 10 cosmological zoom simulations of typical present-day disc galaxies. Modelling the number density and initial properties of star clusters associated with each stellar particle, these simulations indicate that the physics governing star cluster formation result in cluster formation histories deviating markedly from the formation histories of field stars (Reina-Campos et al. 2017). Like Renaud et al.~(2017), the E-MOSAICS simulations compute the tidal forces acting upon star clusters throughout the simulation, but in addition use these measurements to estimate the rate at which star clusters are tidally disrupted, in a fashion similar to the scheme of Prieto \& Gnedin (2008). 

Cosmological zoom simulations likely represent the optimal compromise between the competing requirements of detail and diversity. Although they typically focus on the evolution only of an individual galaxy, they model the cosmic environment experienced by all of the star clusters that form within its progenitors. Having a relatively small memory footprint, multiple zooms can often be run with relatively modest computational resources, enabling diverse samples of galaxies to be assembled. Recent zoom simulations of typical disc galaxies evolved to $z=0$ have realised force resolutions of $\sim 10-100$ pc (Agertz \& Kravtsov 2015, 2016; Hopkins et al. 2014; Wetzel et al. 2016), whilst zoom simulations of dwarf galaxies have achieved $\sim$ pc force resolution (e.g. Read et al. 2016; Kim et al. 2016). Such resolution is clearly still some way from that necessary to numerically integrate the internal dynamics of star clusters, but it is adequate to move beyond the widely-used phenomenological treatments of star formation that are calibrated to reproduce the observed Schmidt (or Schmidt-Kennicutt) relation (see e.g. Kennicutt 1998), which relates the gas density (or projected gas density) and star formation rate (or projected star formation rate) integrated over galaxies or star-forming regions. For example, Semenov et al. (2016), building on work by Schmidt et al. (2014), compute the local star formation efficiency per free fall on $\sim 10$ pc scales using subgrid turbulence models, the latter being calibrated against high resolution simulations of star formation in the super-sonic MHD simulations of Padoan et al. (2012). This approach allows for a non-universal star formation efficiency in GMCs, as is observed (e.g. Murray 2011), and may be important for realising the very high efficiency of star formation in the densest peaks of the ISM.

Similarly, at such high resolution, it becomes necessary to move beyond stellar feedback treatments that instantaneously inject energy and/or momentum from an entire simple stellar population. Stellar evolution codes can be used to predict the injection rate of energy and momentum from radiation, stellar winds and SNe over the lifetime of a population (Agertz et al. 2013; Hopkins et al. 2014). Moreover, high resolution simulations are beginning to offer meaningful predictions of the coupling efficiency of SNe in turbulent media (e.g. Creasey, Theuns \& Bower 2013; Kim \& Ostriker 2015; Martizzi et al. 2015) which, in principle, can be parametrised and adopted as subgrid models in cosmological simulations (Hopkins et al. 2014; Agertz \& Kravtsov 2015, 2016; Grisdale et al. 2017). 

The dynamic range required to model star formation explicitly in a broad cosmological context will remain beyond the scope of state-of-the-art computational resources for many years to come, and hybrid approaches marrying numerical and (semi-)analytic  techniques will likely remain the most profitable line of enquiry. By pushing the interface between the two components to ever smaller scales, the influence of restrictive approximations within the latter can be minimised and, in principle, the predictive power of the models increased. Pursuit of this route, however, requires that two key challenges be addressed. Firstly, numerical astrophysicists must remain able to capitalise on the increasing capacity of high performance computing facilities, which are increasingly delivering increased computing power via greater parallelism (i.e. more processors, and more cores per processor) rather than greater processor clock speeds. Secondly, the more detailed physics governing the life cycle of interstellar gas, such as molecular chemistry (Kim et al. 2017) and the formation of dust grains, radiation transport from local sources within the ISM, magneto-hydrodynamical effects and the transport of cosmic rays, must be incorporated into hydrodynamic simulations in a realistic fashion. The lack of convergence in the implementation of relatively simple phenomenological feedback schemes, highlighted by the study of Scannapieco et al. (2012), serves as a reminder that this challenging step is likely to remain frontier science for some time to come.







\section{Conclusion}
Globular clusters (GCs) continue to be a topic of active research with many fundamental questions remaining. These include: How and when did GCs form? Where did they form and how did they assemble into today's GC systems? What was their initial mass function? Answers to these questions will come from both observations and simulations.

Most GCs of the Milky Way, and those around external galaxies,  are very old. Although indications are that the metal-rich (MR) GCs are $\sim$1-1.5 Gyr younger than the metal-poor (MP) ones, it is still difficult to rule out conclusively coeval ages for the two subpopulations.
Absolute ages place the oldest GCs at around 12.5 Gyr but uncertainties extend their ages well into the epoch of reionisation (z $>$ 6).

Near-term prospects for making progress with absolute ages for resolved GCs include obtaining near-IR colour magnitude diagrams that exploit the age sensitivity of the main sequence turnoff and a feature associated with H$_2$ opacity. For unresolved, but nearby, GCs measuring relative ages with the next generation of wide-field, multiplexing spectrographs will be very challenging with 8-10m telescopes. Progress will probably have to wait for large numbers of GCs to be observed with the 20-40m class telescopes and will need to be accompanied by 
a better understanding of theoretical integrated-light age diagnostics. 
An exciting new avenue, that has recently been opened up by HST, and will no doubt be followed-up by JWST, is the direct observation of proto-GC candidates at high redshift. Some fraction of these objects may have survived destruction processes to comprise the present-day population of GCs.

Multiple populations (within an individual GC) appear to be a common feature in old GCs -- indeed, they may be a defining feature of a GC. Recently, all the hallmarks of multiple populations have been found in compact, massive star clusters of age $\sim$ 2 Gyr. Thus the conditions necessary for multiple populations exist in the local universe. 
These young GCs, forming at z $\sim$ 0.17, suggest a common single channel for GC formation. 

Due to internally and externally-driven destruction processes GC systems formed with much more mass than is observed today. 
The commonly discussed formation models for multiple populations require that all individual GCs were at least ten times more massive than they are currently in order to solve the mass budget problem (i.e. to have enough material processed through a 1st generation of stars to form a 2nd generation of stars that matches the present enriched fraction ratio).  However, this universal amount of mass loss is not expected from models of GC evolution, instead it comes from fine tuning the initial conditions in order to match the observed population ratios. Relatively extreme initial conditions must be assumed, and present day GC population properties are not consistent with these assumptions. 
For the Milky Way, assuming a power-law mass function,  the initial mass of the entire GC system has been estimated to be 10-60$\times$ that of its current mass (few $\sim$ 10$^7$ M$_{\odot}$). For dwarf galaxies, such large inferred mass losses are in tension with the large fraction of metal-poor stars in GCs.
The initial shape of the mass function for old GCs is currently unknown -- it is still an open question as to whether the initial mass function resembled a power-law or had a preferred mass scale. 
However, for massive galaxies a power-law initial mass function is currently favoured. 
These issues remain areas of active research.

The marked influence of environment on the evolution of GCs requires that self-consistent models consider GCs within the full cosmological context of their host galaxy. Clearly, such models are also a necessity when seeking to explore and interpret the observed correlations between GCs and their hosts. The dynamic range posed by this necessity in terms of the mass, length and timescales -- particularly when modelling populations of GCs -- presents a major challenge to the modelling community. Moreover, the goalposts are likely to move again in the near future, as the promise of JWST is realised. However, a number of ambitious, complementary modelling approaches have recently emerged, which have demonstrated plausible origins for key present-day GC-galaxy scaling relations. These recent successes should foster genuine optimism for the discovery potential of the forthcoming generation of cosmological simulations of globular cluster systems.

\vskip6pt

\enlargethispage{20pt}

\ethics{Not applicable.}

\dataccess{
This article has made use of NASA's Astrophysics Data System (ADS). }

\aucontribute{DF contributed to the sections on when and where GCs form. AF contributed to the section on where GCs form. JMDK contributed to the sections on how GCs form and numerical approaches. OG contributed to the sections on how GCs form, numerical approaches, and modelling developments. MT contributed to the section on how GCs form. SL contributed to the sections on GC initial masses and ancient vs recently formed GCs. MG contributed to the section on GC initial masses. NB contributed to the sections on when GCs form and ancient vs recently formed GCs. RAC, OA, JP and SP contributed to the section on modelling prospects. All authors discussed the results, commented on the manuscript, and give final approval for publication.}

\competing{
We have no competing interests.}

\funding{All authors thank the Royal Society for their support of the workshop that helped to generate this article. NB, MG and RAC are Royal Society University Research Fellows. NB and JP are partially funded through a European Research Council (ERC) Consolidator grant (ERC-CoG-646928, Multi-Pop). JMDK gratefully acknowledges funding from the German Research Foundation (DFG) in the form of an Emmy Noether Research Group (grant number KR4801/1-1), from the ERC under the European Union's Horizon 2020 research and innovation programme via the ERC Starting Grant MUSTANG (grant agreement number 714907), and from Sonderforschungsbereich SFB 881 ``The Milky Way System'' (subproject P1) of the DFG. DF acknowledges financial support from the Australian Research Council. SP acknowledges support from the European Research Council under the European Union's Seventh Framework Programme (FP7/2007- 2013)/ERC Grant agreement 278594-GasAroundGalaxies.  OA acknowledges support from the Swedish Research Council (grant 2014-5791) and the Knut and Alice Wallenberg Foundation.}

\ack{
We acknowledge the contributions of the other attendees at the Chicheley Hall workshop to the overall discussion. }



\end{document}